\documentclass[aps,prd,amsfonts,amssymb, amsmath, showpacs, showkeys, a4paper, nofootinbib, superscriptaddress, 11pt, raggedbottom]{revtex4-2}

\usepackage{braket}
\usepackage{bm}
\usepackage{graphicx}
\usepackage{slashed}
\usepackage{subfig}
\usepackage{mathrsfs}
\usepackage{color}
\usepackage{mathtools}
\usepackage[normalem]{ulem}
\usepackage{simplewick}
\usepackage{soul}

\usepackage{comment}

\allowdisplaybreaks

\begin{document}

\title{\Large Fock state probability changes in open quantum systems}

\author{Clare Burrage}
\email{clare.burrage@nottingham.ac.uk}
\affiliation{School of Physics and Astronomy, University of Nottingham,
University Park, Nottingham NG7 2RD, UK}

\author{Christian K\"{a}ding}
\email{christian.kaeding@tuwien.ac.at}
\affiliation{Atominstitut, Technische Universit\"at Wien, Stadionallee 2, 1020 Vienna, Austria}

\begin{abstract}
Open quantum systems are powerful effective descriptions of quantum systems interacting with their environments. Studying changes of Fock state probabilities can be intricate in this context since the prevailing description of open quantum dynamics is by master equations of the systems' reduced density matrices, which usually requires finding solutions for a set of complicated coupled differential equations. In this article, we show that such problems can be circumvented by employing a recently developed path integral-based method for directly computing reduced density matrices in scalar quantum field theory. For this purpose, we consider a real scalar field $\phi$ as an open system interacting via a $\lambda \chi^2\phi^2$-term with an environment comprising another real scalar field $\chi$ that has a finite temperature. In particular, we investigate how the probabilities for observing the vacuum or two-particle states change over time if there were initial correlations of these Fock states. Subsequently, we apply our resulting expressions to a neutrino toy model. We show that, within our model, lighter neutrino masses would lead to a stronger distortion of the observable number of particles due to the interaction with the environment after the initial production process.
\end{abstract}


\maketitle



\section{Introduction}

Realistic quantum systems must usually be described as open, which means that they are interacting with other systems, so-called environments, whose degrees of freedom are traced out in order to obtain an effective description within the theory of open quantum systems \cite{Breuer2002}. Open quantum systems regularly find applications in many areas of fundamental physics, including quantum field theory \cite{Calzetta2008,Koksma2010,Koksma2011,Sieberer2016,Marino2016,Baidya2017,Burrage:2018pyg,Burrage2019,Banerjee:2020ljo,Nagy2020,Banerjee:2021lqu,Jana2021,Fogedby2022,Kading:2022jjl,Cao:2023syu,Bowen:2024emo,Reyes-Osorio:2024chg,Keefe:2024cia,Fahn:2024fgc,Kashiwagi:2024fuy,Salcedo:2024nex} or cosmology \cite{Lombardo1,Lombardo2,Lombardo3,Boyanovsky1,Boyanovsky2,Boyanovsky3,Boyanovsky4,Burgess2015,Hollowood,Shandera:2017qkg,Choudhury:2018rjl,Bohra:2019wxu,Akhtar:2019qdn,Binder2021,Brahma:2021mng,Cao:2022kjn,Brahma2022,Colas:2022hlq,Colas:2022kfu,Colas:2023wxa,Kading:2023mdk,Bhattacharyya:2024duw,Colas:2024xjy,Burgess:2024eng,Salcedo:2024smn,Colas:2024lse,Colas:2024ysu,Brahma:2024yor,Kading:2024jqe,Brahma:2024ycc,Burgess:2024heo,Lau:2024mqm}. Mathematically, open quantum systems are often described by reduced density matrices, which are obtained by tracing out the environmental degrees of freedom from the density matrix that describes the closed combined system. The time evolution of reduced density matrices is determined by quantum master equations, which are usually intricate, if not impossible to solve without approximations. Refs.~\cite{Burrage:2018pyg,Burrage2019} presented a powerful method for deriving non-Markovian quantum master equations in quantum field theory,  derived from first principles and  based on the Schwinger-Keldysh closed time-path formalism \cite{Schwinger,Keldysh}. This method already found its first application, to the gravitationally induced decoherence of a scalar field, in Ref.~\cite{Fahn:2024fgc}. In order to avoid having to solve complicated master equations, Ref.~\cite{Kading:2022jjl} extended the method from Refs.~\cite{Burrage:2018pyg,Burrage2019} and found a way of circumventing quantum master equations by instead directly computing reduced density matrices. This new method is also first principle-based, describes non-Markovian dynamics and makes  very few assumptions. It has already been applied to particle phenomenology in Refs.~\cite{Kading:2023mdk,Kading:2024jqe} and  besides being useful for open systems, this method was modified to also be applicable to closed systems; see Ref.~\cite{Kading:2022hhc}.

The effect of an environment on the propagation of particles has been widely studied for many different types of environment and particles. For example,  propagation through a dense medium can lead to changes in the effective masses of neutrinos \cite{ParticleDataGroup:2024cfk}. When thinking of the propagating particles as quantum states, the environment can lead to the loss of coherence of the propagating state(s); see, for example, Ref.~\cite{Breuer:2007juk}. This has been widely studied in the language of open quantum systems. However, other effects of the environment on the evolution of the probability density matrix remain to be explored. 

In this work, we will consider a simplified setup with one scalar field for our `system' and another scalar field for the `environment'. 
 However, the motivation for considering the effects of the environment on a quantum state arises because cosmology and astrophysics are expected to provide such environments that couple to the particles that propagate through our universe, for example, a background of gravitational waves or dark matter fields. One common system, for which the effects of the environment are vitally important to understand, are neutrinos, where the effects of the environment on the propagation can lead to changes in how we understand oscillation between different flavours \cite{ParticleDataGroup:2024cfk}.  This is because neutrino oscillations are a quantum superposition effect. The most general open effective field theory for electromagnetism in a medium has also recently been constructed \cite{Salcedo:2024nex}.

 The effects of the environment on the propagation of neutrinos have been considered, in the language of open quantum systems, previously \cite{Stankevich:2024xyc,Banerjee:2024gtn}. Systems that have been considered include looking at the effects of linearised gravity~\cite{Domi:2024ypm}, squeezed gravitational waves~\cite{Sharifian:2023jem}   the impact of virtual black holes  ~\cite{Stuttard:2020qfv} or in quantum gravity models which give rise to space-time foam \cite{Hawking:1982dj} (more generally, gravitational induced decoherence was considered in Ref.~\cite{Hsiang:2024qou}). However the main focus of these works is on the decoherence of the neutrino states.  Reviews of gravitational decoherence can be found in Refs.~\cite{Anastopoulos:2021jdz,Bassi:2017szd}.
 In Ref.~\cite{Ribeiro:2024yue}, the authors develop a `Collisional Approach for Open Neutrino Systems' which derives a master equation for the propagation of neutrinos coupled to an environment, and derives new bounds for the interaction of neutrinos with an ultra-light dark matter background.  They discuss effects beyond just decoherence, including those that change the number of neutrinos (e.g. neutrino decay) but to solve the resultant infinite coupled set of master of equations they have to make a number of assumptions.

Treatment of a background of ultra-light dark matter as an environment for quantum systems has been discussed in Ref.~\cite{Bernal:2024hcc}. Quantum master equations for ultra-light dark matter have been derived in Ref.~\cite{Cao:2022kjn}, leading to the suggestion that  the dark matter is formed of two components, a cold condensate, and a hot component that arises due to thermalisation with the cosmological thermal bath. The effects of an ultra-light dark matter background \cite{Antypas:2022asj,Hui:2021tkt,Ferreira:2020fam} on the propagation of neutrinos have also been considered \cite{Chun:2021ief,Berlin:2016woy,Krnjaic:2017zlz,Brdar:2017kbt}. Indeed, it has been suggested that  interaction with this background could account for the observed neutrino masses~\cite{Sen:2023uga}, or give rise to redshift dependent masses \cite{Martinez-Mirave:2024dmw}.

The impacts of thermal environments on the propagation and oscillations of neutrinos have also been considered previously, for example, in Ref.~\cite{Stodolsky:1986dx}, but never, as far as we are aware, using an open quantum system approach. Though, the effects of a thermal environment on an open system have been considered more generally; see, for example, Ref.~\cite{Boyanovsky:2015xoa}.

 In this work, we show that working at the level of the density matrix, rather than the master equation, one can more easily study the effects of the environment on the quantum evolution of a system.  In particular, this does not just provide information about decoherence, but also about possible changes in the probability of finding a certain number of particles present. This builds on previous work by one of us \cite{Kading:2022jjl}
 which introduced this formalism. 
 
 We work in a simplified model  with two scalar fields interacting, one system one environment.  The interaction is through a `portal' term \cite{Beacham:2019nyx} which is controlled by a dimensionless parameter $\lambda$. As we only work with scalar fields, our results will not be directly applicable to neutrino oscillations, but we hope this simpler example shows the utility of our approach, and indicates the potential for applying this to neutrino oscillations as well as other open quantum systems.
We are able to consider the effects of both  quantum and thermal fluctuations in the environment.  However,  after introducing counter terms (regularisation), we will find that we are left with only the thermal correction.  This is independent of the masses of the scalars, although
all parameters we choose must comply with perturbativity.

In Section \ref{sec:Prob}, we introduce the model of two coupled scalar fields, and show how to compute the probability of finding the system in the vacuum or a two particle state after a period of free evolution. In Section \ref{sec:Toy}, we make our results quantitative, by choosing values for our parameters illustrative of a neutrino propagating in a cosmological background. We conclude in Section \ref{sec:Conclusion}. 


\section{Probabilities for vacuum and two-particle states}
\label{sec:Prob}

In this section, we introduce the model that we will be working with throughout the article and compute the probabilities for finding the considered system at an arbitrary time $t$ in either a vacuum or two-particle state. As our model, we consider a real scalar field $\phi$ (the system) with mass $M$ that is interacting with an environment comprising another real scalar field $\chi$ with mass $m$ and temperature $T$. Consequently, the free actions of the two scalar fields are given by
\begin{eqnarray}
S_\phi[\phi] &=& \int_x \left[ -\frac{1}{2}(\partial\phi)^2 - \frac{1}{2}M^2\phi^2 \right]
~,~~~
S_\chi[\chi] = \int_x \left[ -\frac{1}{2}(\partial\chi)^2 - \frac{1}{2}m^2\chi^2 \right]~,
\end{eqnarray}
and we choose the interaction
\begin{eqnarray}
\label{eq:Interaction}
S_{\text{int}}[\phi,\chi] &=& \int_{x\in\Omega_t} \left[ -\lambda \chi^2\phi^2 \right]~,
\end{eqnarray}
where $\lambda \ll 1$ is a dimensionless coupling constant and, since we discuss only a finite time interval, $\Omega_t := [0,t] \times \mathbb{R}^3$. We assume that, at the initial time $t=0$, the system is in a superposition of zero and two $\phi$-particles, such that the initial system (reduced) density matrix can be expanded in a Fock basis as \cite{Kading:2022jjl}
\begin{eqnarray}
\hat{\rho}(0) 
&=& 
\rho_{0;0}(;;0) \ket{0}\bra{0} 
+ 
\frac{1}{2} \int d\Pi_{\mathbf{k}_1} d\Pi_{\mathbf{k}_2}
\rho_{2;0}(\mathbf{k}_1,\mathbf{k}_2;;0) \ket{\mathbf{k}_1,\mathbf{k}_2}\bra{0}
\nonumber
\\
&&
\phantom{\rho_{0;0}(;;0) \ket{0}\bra{0} }
+
\frac{1}{2} \int d\Pi_{\mathbf{k}'_1} d\Pi_{\mathbf{k}'_2}
\rho_{0;2}(;\mathbf{k}'_1,\mathbf{k}'_2;0) \ket{0}\bra{\mathbf{k}'_1,\mathbf{k}'_2}
\nonumber
\\
&&
+
\frac{1}{4} \int d\Pi_{\mathbf{k}_1} d\Pi_{\mathbf{k}_2} d\Pi_{\mathbf{k}'_1} d\Pi_{\mathbf{k}'_2}
\rho_{2;2}(\mathbf{k}_1,\mathbf{k}_2;\mathbf{k}'_1,\mathbf{k}'_2;0) \ket{\mathbf{k}_1,\mathbf{k}_2}\bra{\mathbf{k}'_1,\mathbf{k}'_2}~,
\end{eqnarray}
where 
\begin{eqnarray}
\int d\Pi_{\mathbf{k}} &:=& \int_{\mathbf{k}} \frac{1}{2E^\phi_{\mathbf{k}}}~,~~~
\int_{\mathbf{k}} := \int \frac{d^3k}{(2\pi)^3}~,
\end{eqnarray}
and $E^\phi_\mathbf{k} := \sqrt{\mathbf{k}^2 + M^2}$ is the energy of the field $\phi$ with momentum $\mathbf{k}$.
Here, the density matrix elements $\rho_{0;0}$, $\rho_{2;0}$, and $\rho_{2;2}$ describe the vacuum state, the correlation between vacuum and two-particle state, and the two-particle state, respectively. Note that, as usual, $\rho_{2;0}(\mathbf{k}_1,\mathbf{k}_2;;0) = \rho^\ast_{0;2}(;\mathbf{k}_1,\mathbf{k}_2;0)$ and $\rho_{0;0},\rho_{2;2} \in \mathbb{R}$.

Generally, the probability of finding the system in an $N$-particle state at time $t$ is given by
\begin{eqnarray}
    P_N &=& \mathrm{Tr}[\hat{\mathcal{P}}_N \hat{\rho}(t)]~,
\end{eqnarray}
where $\hat{\mathcal{P}}_N$ is the projector into the $N$-particle subspace. Since we are interested in determining the probabilities of observing the system in the vacuum state $\ket{0}$ or in a two-particle state, we must compute
\begin{eqnarray}
\label{eq:P0t0}
    P_0(t) &=& \mathrm{Tr}[\ket{0}\bra{0} \hat{\rho}(t)] = \rho_{0;0}(;;t)~,
\end{eqnarray}
and 
\begin{eqnarray}
\label{eq:P2t0}
    P_2(t) &=& \mathrm{Tr}\left[\frac{1}{2} \int d\Pi_{\mathbf{p}_1} d\Pi_{\mathbf{p}_2} 
 \ket{\mathbf{p}_1,\mathbf{p}_2}\bra{\mathbf{p}_1,\mathbf{p}_2} \hat{\rho}(t)\right]
   =
   \frac{1}{2} \int d\Pi_{\mathbf{p}_1} d\Pi_{\mathbf{p}_2} 
\rho_{2;2}(\mathbf{p}_1,\mathbf{p}_2;\mathbf{p}_1,\mathbf{p}_2;t)~.
\end{eqnarray}
If we assume that there has been no interaction between system and environment before the initial time, then from unitarity follows that
\begin{eqnarray}
\label{eq:unitarity}
    1 &=& P_0(0) + P_2(0) = \rho_{0;0}(;;0) + \frac{1}{2} \int d\Pi_{\mathbf{k}_1} d\Pi_{\mathbf{k}_2} 
\rho_{2;2}(\mathbf{k}_1,\mathbf{k}_2;\mathbf{k}_1,\mathbf{k}_2;0)~.
\end{eqnarray}
Note that, generally, 
\begin{eqnarray}
    1 &\geq& P_0(t) + P_2(t) = \rho_{0;0}(;;t) + \frac{1}{2} \int d\Pi_{\mathbf{p}_1} d\Pi_{\mathbf{p}_2} 
\rho_{2;2}(\mathbf{p}_1,\mathbf{p}_2;\mathbf{p}_1,\mathbf{p}_2;t)
\end{eqnarray}
since there is also a certain probability that $(N > 2)$-particle states will be occupied within the time interval $[0,t]$. However, since we are restricting our discussion to first order in $\lambda$, such states are not of relevance for us and, at the considered order, there are no non-unitarian effects.

In order to obtain the probabilities in Eqs.~(\ref{eq:P0t0}) and (\ref{eq:P2t0}), we must first compute the required reduced density matrix elements. For this, we employ the method presented in Ref.~\cite{Kading:2022jjl}, which is first principle-based, can describe non-Markovian dynamics, only requires the assumption of no initial correlations between system and environment, and enables us to directly compute reduced density matrix elements without having to solve quantum master equations. In the context of quantum field theory, this method is, as of yet, the only one that allows us to directly compute reduced density matrices in a phenomenologically relevant momentum basis. Furthermore, what makes this method different from others in the literature is the fact that, after tracing out the environmental degrees of freedom, exact expressions for the reduced density matrix elements are obtained. Only for evaluating these expressions, perturbation theory is employed, and not before tracing out the environment. Since we will be restricting our discussion to terms of first order in $\lambda$, we do not expect to see non-Markovian dynamics here. Nevertheless, from our subsequent computations, it will be clear that directly computing the density matrix elements is more easily applicable for describing particle number-changing processes in open systems than the usual master equation approach. We have to evaluate the following (exact) expressions:
\begin{eqnarray}
\label{eq:VacDens0}
\rho_{0;0}(;;t)
&=& 
\rho_{0;0}(;;0) 
\int\mathcal{D}\phi^{\pm} e^{\mathrm{i}\widehat{S}_{\phi}[\phi]}\widehat{\mathcal{F}}[\phi;t]
\nonumber
\\
&&
-\frac{1}{2}
\lim_{\substack{y_1^{0},y_2^0\,\to\, 0^-}}
\int d\Pi_{\mathbf{k}_1} d\Pi_{\mathbf{k}_2}\rho_{2;0}(\mathbf{k}_1,\mathbf{k}_2;;0) 
\nonumber
\\
&&
\times 
\int_{\mathbf{y}_1\mathbf{y}_2} e^{\mathrm{i}(\mathbf{k}_1\mathbf{y}_1+\mathbf{k}_2\mathbf{y}_2)}
\partial_{y_1^0,E^\phi_{\mathbf{k}_1}}^*\partial_{y_2^0,E^\phi_{\mathbf{k}_2}}^*
\int\mathcal{D}\phi^{\pm} e^{\mathrm{i}\widehat{S}_{\phi}[\phi]}\widehat{\mathcal{F}}[\phi;t]\phi^{+}_{y_1}\phi^{+}_{y_2}
\nonumber
\\
&&
- \frac{1}{2}
\lim_{\substack{y_1^{0\prime},y_2^{0\prime}\,\to\, 0^-}}
\int d\Pi_{\mathbf{k}'_1} d\Pi_{\mathbf{k}'_2}\rho_{0;2}(;\mathbf{k}'_1,\mathbf{k}'_2;0) 
\nonumber
\\
&&
\times 
\int_{\mathbf{y}_1\mathbf{y}_2} e^{-\mathrm{i}(\mathbf{k}'_1\mathbf{y}'_1+\mathbf{k}'_2\mathbf{y}'_2)}
\partial_{y_1^{0\prime},E^\phi_{\mathbf{k}'_1}}\partial_{y_2^{0\prime},E^\phi_{\mathbf{k}'_2}}
\int\mathcal{D}\phi^{\pm} e^{\mathrm{i}\widehat{S}_{\phi}[\phi]}\widehat{\mathcal{F}}[\phi;t]\phi^{-}_{y_1'}\phi^{-}_{y_2'}
~,
\end{eqnarray}
and
\begin{eqnarray}
\label{eq:2Dens0}
\rho_{2;2}(\mathbf{p}_1,\mathbf{p}_2;\mathbf{p}'_1,\mathbf{p}'_2;t)
&=& 
\frac{1}{4}
\lim_{\substack{x_1^{0(\prime)},x_2^{0(\prime)}\,\to\, t^{+}\\y_1^{0(\prime)},y_2^{0(\prime)}\,\to\, 0^-}}
\int d\Pi_{\mathbf{k}_1} d\Pi_{\mathbf{k}_2}d\Pi_{\mathbf{k}'_1}d\Pi_{\mathbf{k}'_2} \rho_{2;2}(\mathbf{k}_1,\mathbf{k}_2;\mathbf{k}'_1,\mathbf{k}'_2;0) 
\nonumber
\\
&&
\times 
\int_{\mathbf{x}_1\mathbf{x}_2\mathbf{x}'_1\mathbf{x}'_2\mathbf{y}_1\mathbf{y}_2\mathbf{y}'_1\mathbf{y}'_2} e^{-\mathrm{i}(\mathbf{p}_1\mathbf{x}_1 + \mathbf{p}_2\mathbf{x}_2-\mathbf{p}'_1 \mathbf{x}'_1-\mathbf{p}'_2\mathbf{x}'_2)+\mathrm{i}(\mathbf{k}_1\mathbf{y}_1+\mathbf{k}_2\mathbf{y}_2-\mathbf{k}'_1\mathbf{y}'_1 -\mathbf{k}'_2\mathbf{k}'_2)}
\nonumber
\\
&&
\times 
\partial_{x_1^0,E^\phi_{\mathbf{p}_1}} 
\partial_{x_2^0,E^\phi_{\mathbf{p}_2}}\partial_{x_1^{0\prime},E^\phi_{\mathbf{p}'_1}}^*\partial_{x_2^{0\prime},E^\phi_{\mathbf{p}'_2}}^*
\partial_{y_1^0,E^\phi_{\mathbf{k}_1}}^*\partial_{y_2^0,E^\phi_{\mathbf{k}_2}}^*\partial_{y_1^{0\prime},E^\phi_{\mathbf{k}'_1}}\partial_{y_2^{0\prime},E^\phi_{\mathbf{k}'_2}}
\nonumber
\\
&&
\times 
\int\mathcal{D}\phi^{\pm} e^{\mathrm{i}\widehat{S}_{\phi}[\phi]}\phi^+_{x_1}\phi^+_{x_2}\phi^-_{x'_1}\phi^-_{x'_2}\widehat{\mathcal{F}}[\phi;t]\phi^{+}_{y_1}\phi^{+}_{y_2}\phi^{-}_{y'_1}\phi^{-}_{y'_2}
\nonumber
\\
&&
-\frac{1}{2}
\lim_{\substack{x_1^{0(\prime)},x_2^{0(\prime)}\,\to\, t^{+}\\y_1^{0},y_2^0\,\to\, 0^-}}
\int d\Pi_{\mathbf{k}_1} d\Pi_{\mathbf{k}_2}\rho_{2;0}(\mathbf{k}_1,\mathbf{k}_2;;0) 
\nonumber
\\
&&
\times 
\int_{\mathbf{x}_1\mathbf{x}_2\mathbf{x}'_1\mathbf{x}'_2\mathbf{y}_1\mathbf{y}_2} e^{-\mathrm{i}(\mathbf{p}_1\mathbf{x}_1 + \mathbf{p}_2\mathbf{x}_2-\mathbf{p}'_1 \mathbf{x}'_1-\mathbf{p}'_2\mathbf{x}'_2)+\mathrm{i}(\mathbf{k}_1\mathbf{y}_1+\mathbf{k}_2\mathbf{y}_2)}
\nonumber
\\
&&
\times 
\partial_{x_1^0,E^\phi_{\mathbf{p}_1}} 
\partial_{x_2^0,E^\phi_{\mathbf{p}_2}}\partial_{x_1^{0\prime},E^\phi_{\mathbf{p}'_1}}^*\partial_{x_2^{0\prime},E^\phi_{\mathbf{p}'_2}}^*
\partial_{y_1^0,E^\phi_{\mathbf{k}_1}}^*\partial_{y_2^0,E^\phi_{\mathbf{k}_2}}^*
\nonumber
\\
&&
\times 
\int\mathcal{D}\phi^{\pm} e^{\mathrm{i}\widehat{S}_{\phi}[\phi]}\phi^+_{x_1}\phi^+_{x_2}\phi^-_{x'_1}\phi^-_{x'_2}\widehat{\mathcal{F}}[\phi;t]\phi^{+}_{y_1}\phi^{+}_{y_2}
\nonumber
\\
&&
-\frac{1}{2}
\lim_{\substack{x_1^{0(\prime)},x_2^{0(\prime)}\,\to\, t^{+}\\y_1^{0\prime},y_2^{0\prime}\,\to\, 0^-}}
\int d\Pi_{\mathbf{k}'_1} d\Pi_{\mathbf{k}'_2}\rho_{0;2}(;\mathbf{k}'_1,\mathbf{k}'_2;0) 
\nonumber
\\
&&
\times 
\int_{\mathbf{x}_1\mathbf{x}_2\mathbf{x}'_1\mathbf{x}'_2\mathbf{y}_1\mathbf{y}_2} e^{-\mathrm{i}(\mathbf{p}_1\mathbf{x}_1 + \mathbf{p}_2\mathbf{x}_2-\mathbf{p}'_1 \mathbf{x}'_1-\mathbf{p}'_2\mathbf{x}'_2)-\mathrm{i}(\mathbf{k}'_1\mathbf{y}'_1+\mathbf{k}'_2\mathbf{y}'_2)}
\nonumber
\\
&&
\times 
\partial_{x_1^0,E^\phi_{\mathbf{p}_1}} 
\partial_{x_2^0,E^\phi_{\mathbf{p}_2}}\partial_{x_1^{0\prime},E^\phi_{\mathbf{p}'_1}}^*\partial_{x_2^{0\prime},E^\phi_{\mathbf{p}'_2}}^*
\partial_{y_1^{0\prime},E^\phi_{\mathbf{k}'_1}}\partial_{y_2^{0\prime},E^\phi_{\mathbf{k}'_2}}
\nonumber
\\
&&
\times 
\int\mathcal{D}\phi^{\pm} e^{\mathrm{i}\widehat{S}_{\phi}[\phi]}\phi^+_{x_1}\phi^+_{x_2}\phi^-_{x'_1}\phi^-_{x'_2}\widehat{\mathcal{F}}[\phi;t]\phi^{-}_{y_1'}\phi^{-}_{y_2'}~, 
\end{eqnarray}
where $\phi_x := \phi(x)$, $\widehat{S}_{\phi}[\phi] := S_{\phi}[\phi^+] - S_{\phi}[\phi^-]$, the superscripts ${}^+$ and ${}^-$ label the two different branches of the Schwinger-Keldysh closed time path \cite{Schwinger,Keldysh}, $\partial_{t,E^\phi_{\mathbf{p}}} := \overset{\rightarrow}{\partial}_t - \mathrm{i}E^\phi_{\mathbf{p}}$, and $\widehat{\mathcal{F}}[\phi;t]$ is the Feynman-Vernon influence functional \cite{Feynman:1963fq} that can be defined by \cite{Kading:2023mdk}
\begin{eqnarray}
\widehat{\mathcal{F}}[\phi;t]  &:=& \left\langle e^{\mathrm{i}\big\{   S_\text{int}[\phi^+,\chi^+;t]- S_\text{int}[\phi^-,\chi^-;t] \big\}} \right\rangle_\chi 
\end{eqnarray}
with the expectation value
\begin{eqnarray} 
\langle A[\chi^{a}]\rangle_\chi &:=& \int d\chi^{\pm}_t d\chi^{\pm}_0 \delta(\chi_t^+-\chi_t^-)\rho_\chi [\chi^{\pm}_0;0]
\int^{\chi^{\pm}_t}_{\chi^{\pm}_0} \mathcal{D}\chi^{\pm} A[\chi^{a}]e^{\mathrm{i}\left\{S_{\chi}[\chi^+]-S_{\chi}[\chi^-]\right\}}~,
\end{eqnarray} 
in which $\rho_\chi [\chi^{\pm}_0;0]$ denotes the initial density matrix of the environment in a field basis. 
Note that the first line on the right-hand side of Eq.~(\ref{eq:VacDens0}) describes the time evolution of the vacuum density matrix element beginning at the initial time, while the subsequent lines describe how this evolution is modified due to the presence of correlations between zero- and two-particle states at the initial time. For the latter, the Lehmann–Symanzik–Zimmermann-like reduction, introduced in Ref.~\cite{Burrage:2018pyg}, was employed. Eq.~(\ref{eq:2Dens0}) has essentially the same structure for the evolution of the two-particle density matrix.

Expanding the Feynman-Vernon influence functional up to first order in $\lambda$ gives
\begin{eqnarray}
\widehat{\mathcal{F}}[\phi;t] 
&\approx& 1 + \mathrm{i}\sum\limits_{a=\pm} a\langle S_\text{int}[\phi^a,\chi^a;t] \rangle_\chi 
\approx 1 -\mathrm{i}\lambda\sum\limits_{a=\pm} a \int_x \Delta^{\rm F}_{xx}  (\phi_x^a)^2 
~,
\end{eqnarray}
where 
\begin{eqnarray}
    \Delta^{\rm F}_{xx}
&=&  -\mathrm{i} \int_k \left[\frac{1}{k^2+m^2-\mathrm{i}\epsilon} +2\pi \mathrm{i} f(|k^0|) \delta(k^2+m^2) \right]
\end{eqnarray}
is the tadpole Feynman propagator of the environment $\chi$ with
\begin{eqnarray}
    \int_k &:=& \int \frac{d^4k}{(2\pi)^4}
\end{eqnarray}
and 
\begin{eqnarray}
    f(k^0)&:=&\frac{1}{e^{k^0/T}-1} 
\end{eqnarray}
being the Bose-Einstein distribution. Evaluating the path integrals in Eqs.~(\ref{eq:VacDens0}) and (\ref{eq:2Dens0}) gives rise to the system propagators 
\begin{eqnarray}
\contraction{}{\phi}{^+_x}{\phi}\phi^+_x\phi^+_y &=& D^\mathrm{F}_{xy} = - \mathrm{i}\int_k \frac{e^{\mathrm{i}k (x-y)}}{k^2+M^2-i\epsilon}~,~~~
\contraction{}{\phi}{^-_x}{\phi}\phi^-_x\phi^-_y = D^\mathrm{D}_{xy} = + \mathrm{i}\int_k \frac{e^{\mathrm{i}k (x-y)}}{k^2+M^2+i\epsilon}~.
\end{eqnarray}
Note that, for the system, no contractions of two fields labeled with $+$ and $-$ are permitted \cite{Kading:2022jjl}. Consequently, up to first order in $\lambda$, we obtain
\begin{eqnarray}
\rho_{0;0}(;;t)
&\approx& 
\rho_{0;0}(;;0) 
+\Bigg[\mathrm{i}\lambda
\lim_{\substack{y_1^{0},y_2^0\,\to\, 0^-}}
\int d\Pi_{\mathbf{k}_1} d\Pi_{\mathbf{k}_2}
\rho_{2;0}(\mathbf{k}_1,\mathbf{k}_2;;0) 
\int_{\mathbf{y}_1\mathbf{y}_2} e^{\mathrm{i}(\mathbf{k}_1\mathbf{y}_1+\mathbf{k}_2\mathbf{y}_2)}
\partial_{y_1^0,E^\phi_{\mathbf{k}_1}}^*\partial_{y_2^0,E^\phi_{\mathbf{k}_2}}^*
\nonumber
\\
&& 
~~~~~~~~~~~~~~~~~~~~~~~~~~~~~~~~~~~~~~~~~~~~~~~~~~~~~~~~~~~~~~~~~~~~~~~
\times
\int_z
\Delta^{\rm F}_{zz} D^{\rm F}_{zy_1} D^{\rm F}_{zy_2}
+ \mathrm{c.c.}
\Bigg]
\nonumber
\\
&\approx& 
\rho_{0;0}(;;0) 
- \lambda 
\Delta^{\rm F}_{zz} \int 
d\Pi_{\mathbf{q}}
\frac{\sin(E^\phi_{\mathbf{q}}t)}{ (E^\phi_{\mathbf{q}})^2}
\nonumber
\\
&&
~~~~~~~~~~~~~~~~~~~~~~~~
\times
\left\{
\text{Re}[\rho_{2;0}(\mathbf{q},-\mathbf{q};;0)]\sin(E^\phi_{\mathbf{q}}t)
-
\text{Im}[\rho_{2;0}(\mathbf{q},-\mathbf{q};;0)]\cos(E^\phi_{\mathbf{q}}t)
\right\}
~~~~
\end{eqnarray}
and
\begin{eqnarray}
\label{eq:DisConBub}
\rho_{2;2}(\mathbf{p}_1,\mathbf{p}_2;\mathbf{p}'_1,\mathbf{p}'_2;t)
&\approx& 
\lim_{\substack{x_1^{0(\prime)},x_2^{0(\prime)}\,\to\, t^{+}\\y_1^{0(\prime)},y_2^{0(\prime)}\,\to\, 0^-}}
\int d\Pi_{\mathbf{k}_1} d\Pi_{\mathbf{k}_2}d\Pi_{\mathbf{k}'_1}d\Pi_{\mathbf{k}'_2} \rho_{2;2}(\mathbf{k}_1,\mathbf{k}_2;\mathbf{k}'_1,\mathbf{k}'_2;0) 
\nonumber
\\
&&
\times 
\int_{\mathbf{x}_1\mathbf{x}_2\mathbf{x}'_1\mathbf{x}'_2\mathbf{y}_1\mathbf{y}_2\mathbf{y}'_1\mathbf{y}'_2} e^{-\mathrm{i}(\mathbf{p}_1\mathbf{x}_1 + \mathbf{p}_2\mathbf{x}_2-\mathbf{p}'_1 \mathbf{x}'_1-\mathbf{p}'_2\mathbf{x}'_2)+\mathrm{i}(\mathbf{k}_1\mathbf{y}_1+\mathbf{k}_2\mathbf{y}_2-\mathbf{k}'_1\mathbf{y}'_1 -\mathbf{k}'_2\mathbf{k}'_2)}
\nonumber
\\
&&
\times 
\partial_{x_1^0,E^\phi_{\mathbf{p}_1}} 
\partial_{x_2^0,E^\phi_{\mathbf{p}_2}}\partial_{x_1^{0\prime},E^\phi_{\mathbf{p}'_1}}^*\partial_{x_2^{0\prime},E^\phi_{\mathbf{p}'_2}}^*
\partial_{y_1^0,E^\phi_{\mathbf{k}_1}}^*\partial_{y_2^0,E^\phi_{\mathbf{k}_2}}^*\partial_{y_1^{0\prime},E^\phi_{\mathbf{k}'_1}}\partial_{y_2^{0\prime},E^\phi_{\mathbf{k}'_2}}
\nonumber
\\
&&
\times 
\Bigg\{
D^{\rm D}_{x'_1y'_1}D^{\rm D}_{x'_2y'_2}
\Bigg[\frac{1}{2}
D^{\rm F}_{x_1y_1}D^{\rm F}_{x_2y_2}
\nonumber
\\
&&
~~~~~~~~~~~~~~~~~~~~~
-2\mathrm{i}\lambda \int_z \Delta^{\rm F}_{zz}
\Bigg(
 D^{\rm F}_{x_1y_1}D^{\rm F}_{x_2z}D^{\rm F}_{zy_2} 
+ D^{\rm F}_{x_1z}D^{\rm F}_{x_2y_2}D^{\rm F}_{zy_1} 
\Bigg)
\Bigg]
\nonumber
\\
&&
~~~~~
+[(x_1,x_2,y_1,y_2)\longleftrightarrow(x'_1,x'_2,y'_1,y'_2)]^\ast
\Bigg\}
\nonumber
\\
&&
-\Bigg\{
2\mathrm{i}\lambda
\lim_{\substack{x_1^{0(\prime)},x_2^{0(\prime)}\,\to\, t^{+}\\y_1^{0},y_2^0\,\to\, 0^-}}
\int d\Pi_{\mathbf{k}_1} d\Pi_{\mathbf{k}_2}\rho_{2;0}(\mathbf{k}_1,\mathbf{k}_2;;0) 
\nonumber
\\
&&
~~~~
\times 
\int_{\mathbf{x}_1\mathbf{x}_2\mathbf{x}'_1\mathbf{x}'_2\mathbf{y}_1\mathbf{y}_2} e^{-\mathrm{i}(\mathbf{p}_1\mathbf{x}_1 + \mathbf{p}_2\mathbf{x}_2-\mathbf{p}'_1 \mathbf{x}'_1-\mathbf{p}'_2\mathbf{x}'_2)+\mathrm{i}(\mathbf{k}_1\mathbf{y}_1+\mathbf{k}_2\mathbf{y}_2)}
\nonumber
\\
&&
~~~~
\times 
\partial_{x_1^0,E^\phi_{\mathbf{p}_1}} 
\partial_{x_2^0,E^\phi_{\mathbf{p}_2}}\partial_{x_1^{0\prime},E^\phi_{\mathbf{p}'_1}}^*\partial_{x_2^{0\prime},E^\phi_{\mathbf{p}'_2}}^*
\partial_{y_1^0,E^\phi_{\mathbf{k}_1}}^*\partial_{y_2^0,E^\phi_{\mathbf{k}_2}}^*
\nonumber
\\
&&
~~~~
\times 
\int_z \Delta^{\rm F}_{zz} 
D^{\rm F}_{x_1y_1}D^{\rm F}_{x_2y_2} D^{\rm D}_{x'_1z}D^{\rm D}_{x'_2z}
+
[(\mathbf{p}_1,\mathbf{p}_2)\longleftrightarrow(\mathbf{p}'_1,\mathbf{p}'_2)]^\ast
\Bigg\}
\nonumber
\\
&\approx&
\Bigg[
1 - \mathrm{i}\lambda t \Delta^{\rm F}_{zz} \Bigg(
\frac{1}{E^\phi_{\mathbf{p_1}}} + \frac{1}{E^\phi_{\mathbf{p_2}}} - \frac{1}{E^\phi_{\mathbf{p'_1}}} - \frac{1}{E^\phi_{\mathbf{p'_2}}}
\Bigg)
\Bigg]
\nonumber
\\
&&
~~~
\times
e^{-\mathrm{i}(E^\phi_{\mathbf{p}_1} + E^\phi_{\mathbf{p}_2} -E^\phi_{\mathbf{p}'_1} -E^\phi_{\mathbf{p}'_2})t}\rho_{2;2}(\mathbf{p}_1,\mathbf{p}_2;\mathbf{p}'_1,\mathbf{p}'_2;0)
\nonumber
\\
&&
-
\Bigg\{
\lambda
\frac{\Delta^{\rm F}_{zz}}{E^\phi_{\mathbf{p}'_1}} (2\pi)^3 \delta^{(3)}(\mathbf{p}'_1 + \mathbf{p}'_2)
e^{-\mathrm{i}(E^\phi_{\mathbf{p}_1}+E^\phi_{\mathbf{p}_2})t}
\rho_{2;0}(\mathbf{p}_1,\mathbf{p}_2;;0) 
\left( 1- e^{2\mathrm{i}E^\phi_{\mathbf{p}'_1}t} \right)
\nonumber
\\
&&
~~~~
+
[(\mathbf{p}_1,\mathbf{p}_2)\longleftrightarrow(\mathbf{p}'_1,\mathbf{p}'_2)]^\ast
\Bigg\}
~,~~~~~~~~~~~~~~~~~~~~~~~~~~~~~~~~~~~~~~~~~~~~~~~~~~~~~~~
\end{eqnarray}
where we have dropped all disconnected bubble diagrams in Eq.~(\ref{eq:DisConBub}) since they could simply be absorbed by a redefinition of the density matrix elements \cite{Burrage:2018pyg}. Following Refs.~\cite{Burrage:2018pyg,Kading:2023mdk}, we introduce a counter term 
\begin{eqnarray}
\delta S_{\text{int}}[\phi, \chi] &:=&    \ \lambda \int_{x}  \Delta^{F(T=0)}_{xx}  \phi^2  
\end{eqnarray}
in the interaction in Eq.~(\ref{eq:Interaction}) in order to cancel the divergent part 
\begin{eqnarray}
    \Delta^{F(T=0)}_{xx}
&:=&  -\mathrm{i} \int_k \frac{1}{k^2+m^2-\mathrm{i}\epsilon} 
\end{eqnarray}
of the tadpole propagators. Consequently, we are left with 
\begin{eqnarray}
\label{eq:rho0000}
\rho_{0;0}(;;t)
&\approx& 
\rho_{0;0}(;;0) 
- \lambda 
\Delta^{\mathrm{F}(T\neq 0)}_{zz} \int 
d\Pi_{\mathbf{q}}
\frac{\sin(E^\phi_{\mathbf{q}}t)}{ (E^\phi_{\mathbf{q}})^2}
\nonumber
\\
&&
~~~~~~~~~~~~~~~~~~~~~~~~
\times
\left\{
\text{Re}[\rho_{2;0}(\mathbf{q},-\mathbf{q};;0)]\sin(E^\phi_{\mathbf{q}}t)
-
\text{Im}[\rho_{2;0}(\mathbf{q},-\mathbf{q};;0)]\cos(E^\phi_{\mathbf{q}}t)
\right\}~,
~~~~~
\end{eqnarray}
and
\begin{eqnarray}
\label{eq:rho2200}
\rho_{2;2}(\mathbf{p}_1,\mathbf{p}_2;\mathbf{p}'_1,\mathbf{p}'_2;t)
&\approx&
\Bigg[
1 - \mathrm{i}\lambda t \Delta^{\mathrm{F}(T\neq 0)}_{zz} \Bigg(
\frac{1}{E^\phi_{\mathbf{p_1}}} + \frac{1}{E^\phi_{\mathbf{p_2}}} - \frac{1}{E^\phi_{\mathbf{p'_1}}} - \frac{1}{E^\phi_{\mathbf{p'_2}}}
\Bigg)
\Bigg]
\nonumber
\\
&&
~~~
\times
e^{-\mathrm{i}(E^\phi_{\mathbf{p}_1} + E^\phi_{\mathbf{p}_2} -E^\phi_{\mathbf{p}'_1} -E^\phi_{\mathbf{p}'_2})t}\rho_{2;2}(\mathbf{p}_1,\mathbf{p}_2;\mathbf{p}'_1,\mathbf{p}'_2;0)
\nonumber
\\
&&
-
\Bigg\{
\lambda
\frac{\Delta^{\mathrm{F}(T\neq 0)}_{zz}  }{E^\phi_{\mathbf{p}'_1}} (2\pi)^3 \delta^{(3)}(\mathbf{p}'_1 + \mathbf{p}'_2)
e^{-\mathrm{i}(E^\phi_{\mathbf{p}_1}+E^\phi_{\mathbf{p}_2})t}
\rho_{2;0}(\mathbf{p}_1,\mathbf{p}_2;;0) 
\left( 1- e^{2\mathrm{i}E^\phi_{\mathbf{p}'_1}t} \right)
\nonumber
\\
&&
~~~~
+
[(\mathbf{p}_1,\mathbf{p}_2)\longleftrightarrow(\mathbf{p}'_1,\mathbf{p}'_2)]^\ast
\Bigg\}
~,~~~~~~~~~~~~~~~~~~~~~~~~~~~~~~~~~~~~~~~~~~~~~~~~~~~~~~~
\end{eqnarray}
where
\begin{eqnarray}
\Delta^{\mathrm{F}(T\neq 0)}_{zz} 
&=&
\int_k 2\pi f(|k^0|) \delta(k^2+m^2)  \,=\, \frac{T^2}{2\pi^2} \int_{m/T}^\infty d\xi \frac{\sqrt{\xi^2-(\frac{m}{T})^2}}{e^\xi -1}
\end{eqnarray}
is finite in the ultraviolet and reduces to $T^2/12$ in the limit $m \to 0$ \cite{Burrage:2018pyg}. 

After substituting Eqs.~(\ref{eq:rho0000}) and (\ref{eq:rho2200}) into Eqs.~(\ref{eq:P0t0}) and (\ref{eq:P2t0}), we finally find
\begin{eqnarray}
\label{eq:P0t}
    P_0(t) &\approx& \rho_{0;0}(;;0) 
    - \lambda 
\Delta^{\mathrm{F}(T\neq 0)}_{zz} \int d\Pi_{\mathbf{q}} \frac{\sin(E^\phi_{\mathbf{q}}t)}{ (E^\phi_{\mathbf{q}})^2}
\nonumber
\\
&&
~~~~~~~~~~~~~~~~~~~~~~~~~~~~
\times
\left\{
\text{Re}[\rho_{2;0}(\mathbf{q},-\mathbf{q};;0)]\sin(E^\phi_{\mathbf{q}}t)
\right.
\nonumber
\\
&&
~~~~~~~~~~~~~~~~~~~~~~~~~~~~~~~~
\left.
-
\text{Im}[\rho_{2;0}(\mathbf{q},-\mathbf{q};;0)]\cos(E^\phi_{\mathbf{q}}t)
\right\}~,
    \\
\label{eq:P2t}
P_2(t) &\approx& \frac{1}{2} \int d\Pi_{\mathbf{p}_1} d\Pi_{\mathbf{p}_2} 
\rho_{2;2}(\mathbf{p}_1,\mathbf{p}_2;\mathbf{p}_1,\mathbf{p}_2;0)
\nonumber
\\
&&
+ \lambda 
\Delta^{\mathrm{F}(T\neq 0)}_{zz} \int d\Pi_{\mathbf{q}} \frac{\sin(E^\phi_{\mathbf{q}}t)}{ (E^\phi_{\mathbf{q}})^2}
\left\{
\text{Re}[\rho_{2;0}(\mathbf{q},-\mathbf{q};;0)]\sin(E^\phi_{\mathbf{q}}t)
\right.
\nonumber
\\
&&
~~~~~~~~~~~~~~~~~~~~~~~~~~~~~~~~~~~~~
\left.
-
\text{Im}[\rho_{2;0}(\mathbf{q},-\mathbf{q};;0)]\cos(E^\phi_{\mathbf{q}}t)
\right\}~.
\end{eqnarray}
Note that $P_0(t) + P_2(t)  \approx 1$ up to terms of at least second order in $\lambda$. Any deviation of this sum from $1$ is at least of $\mathcal{O}(\lambda^2)$. This illustrates that unitarity is not broken at the considered order in the coupling constant.


\section{Toy model of neutrinos}
\label{sec:Toy}

We now want to study the derived probabilities, in Eqs.~(\ref{eq:P0t}) and (\ref{eq:P2t}), in the context of a toy model similar to the one used in Ref.~\cite{Fahn:2024fgc}, i.e., we use the system $\phi$ as a proxy for an electron neutrino. For this purpose, we choose $M = 0.7\,\mathrm{eV}$ since the mass of $\nu_e$ is smaller than $0.8\,\mathrm{eV}$ \cite{ParticleDataGroup:2024cfk}. The environment $\chi$ is still an arbitrary real scalar field with mass $m$ and temperature $T$. 

\begin{figure} [htbp]
\centering
    \includegraphics[width=11.5cm]{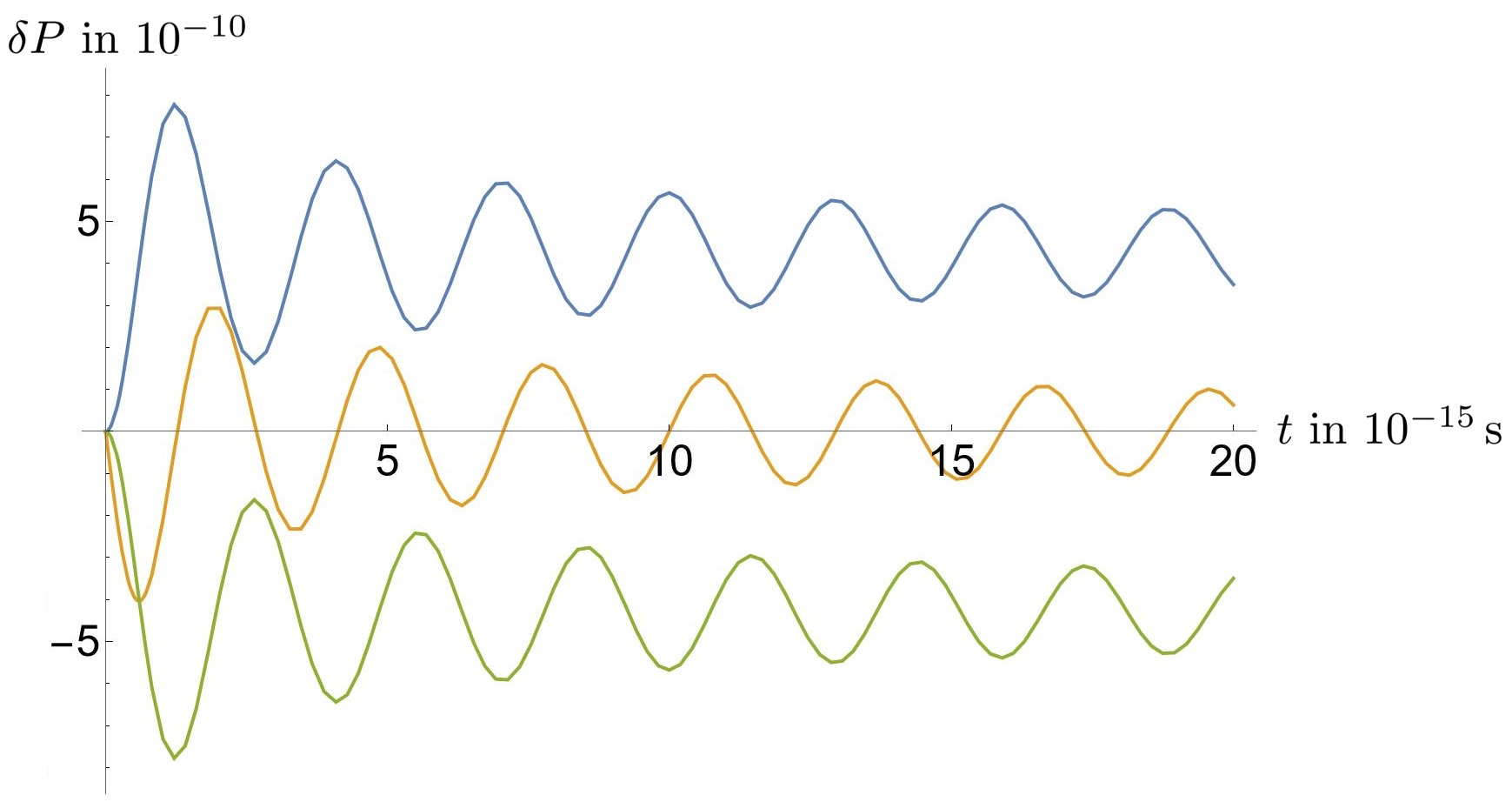}
\caption{Probability deviation $\delta P$ over time $t$ as defined in Eq.~(\ref{eq:delta}) and for the parameters $M= 0.7$\,eV, $T= 2.7$ K, $m = 0$, and $\lambda = 0.1$; the blue, orange, and green lines depict the cases $\theta = 0$, $\theta= \pi/2$, and $\theta = \pi$, respectively.}
\label{fig:First}
\end{figure}

The state of the system at the initial time is highly dependent on the process that produced it. For the considered toy model, we assume an initial state
\begin{eqnarray}
\label{eq:InitialState}
    \ket{\Psi(0)} &=& \sqrt{P_0(0)} \ket{0} 
    + e^{\mathrm{i}\theta}  
    4\sqrt{\pi^3}\sqrt{P_2(0)}\int
    d\Pi_{\mathbf{k}_1} d\Pi_{\mathbf{k}_2} 
    \frac{\sqrt{E^\phi_{\mathbf{k}_1}E^\phi_{\mathbf{k}_2}}}{M |\mathbf{k}_1||\mathbf{k}_2| } e^{-\frac{1}{4M^2} \left( \mathbf{k}_1^2  + \mathbf{k}_2^2 \right)} \ket{\mathbf{k}_1,\mathbf{k}_2}~,
\end{eqnarray}
where $\theta \in \mathbb{R}$ is a phase that we can choose freely and which depends on the specific decay process leading to the two-particle state. Eq.~(\ref{eq:InitialState}) is chosen, so that the probability at time $0$ of finding a neutrino vacuum state or a two-neutrino state is given by $P_0(0)$ or $P_2(0)$, respectively. While we only consider a hypothetical scenario, any production process, left unobserved for a sufficiently long time to lead to a probability of $P_2(0)$ for the two-neutrino state, would, at first order in perturbation theory, lead to such a superposed state. Note that, as is standard in the theory of open quantum systems \cite{Breuer:2007juk}, we must assume that the system and environment were initially not correlated, i.e., the neutrino production process must be shielded from interactions with the environmental scalar field $\chi$. Consequently, we have
\begin{eqnarray}
 \rho_{0;0}(;;0) &=& P_0(0)
 ~,   
 \\
 \rho_{2;0}(\mathbf{k}_1,\mathbf{k}_2;;0) 
 &=&
 e^{\mathrm{i}\theta}8
 \sqrt{\pi^3} \sqrt{P_0(0)P_2(0)}
    \frac{\sqrt{E^\phi_{\mathbf{k}_1}E^\phi_{\mathbf{k}_2}}}{M |\mathbf{k}_1||\mathbf{k}_2| } e^{-\frac{1}{4M^2} \left( \mathbf{k}_1^2  + \mathbf{k}_2^2 \right)}
 ~,
 \\
 \rho_{2;2}(\mathbf{k}_1,\mathbf{k}_2;\mathbf{k}'_1,\mathbf{k}'_2;0)
 &=&
64\pi^3 P_2(0)
 \frac{\sqrt{E^\phi_{\mathbf{k}_1}E^\phi_{\mathbf{k}_2}E^\phi_{\mathbf{k}'_1}E^\phi_{\mathbf{k}'_2}}}{ M^2 |\mathbf{k}_1||\mathbf{k}_2||\mathbf{k}'_1||\mathbf{k}'_2| } e^{-\frac{1}{4M^2} \left( \mathbf{k}_1^2  + \mathbf{k}_2^2 + \mathbf{k}_1^{\prime2}  + \mathbf{k}_2^{\prime2}\right)}
 ~,~~~~~
\end{eqnarray}
which, together with Eq.~(\ref{eq:P2t0}), leads to Eq.~(\ref{eq:unitarity}). Consequently, from Eqs.~(\ref{eq:P0t}) and (\ref{eq:P2t}), we find
\begin{eqnarray}
\label{eq:P0f}
P_0(t) 
&\approx&
P_0(0) 
-  \delta P(t) \sqrt{P_0(0)P_2(0)}
~,~~~
P_2(t) 
\approx
P_2(0)
+
\delta P(t) \sqrt{P_0(0)P_2(0)}
~,
\end{eqnarray}
where
\begin{eqnarray}
\label{eq:delta}
    \delta P(t) &:=& \frac{2\lambda 
\Delta^{\mathrm{F}(T\neq 0)}_{zz}}{\sqrt{\pi} M} \int\limits_0^\infty dq  \frac{\sin(E^\phi_{\mathbf{q}}t)\sin(E^\phi_{\mathbf{q}}t-\theta)}{ (E^\phi_{\mathbf{q}})^2}
e^{-\frac{q^2}{2M^2}  }~.
\end{eqnarray}
In order to quantitatively evaluate the evolution of the probabilities away from their initial values,  Eq.~(\ref{eq:delta}), we first assume $\theta =0$, $T= 2.7$ K, $m = 0$, and $\lambda = 0.1$. These values are chosen to represent a massless scalar, coupled perturbatively to a thermal environment with a cosmological temperature. Doing so, we obtain 
\begin{eqnarray}
\label{eq:deltafirst}
    \delta P(1\text{s}) 
&=&
4 \times 10^{-10}
\end{eqnarray}
for the probability deviation at $t=1$s. The behavior of $\delta P$ as a function of time is depicted as the blue line in Fig.~\ref{fig:First}, where we can see that, after an initial rise, the function oscillates with decreasing magnitude around $+ 4 \times 10^{-10}$. Eqs.~(\ref{eq:P0f}) and (\ref{eq:deltafirst}) show us that, for our particular choice of initial state and parameter values, the probability of observing two neutrinos is increased due to the interaction with the environment. This can be of significance, for example, in situations where a neutrino production process leads to the initial probability of observing two neutrinos being comparable to or smaller than the value of $\delta P$ that we found in Eq.~(\ref{eq:deltafirst}). In such a case, the neutrinos' interaction with the environment would increase the chance of us observing more neutrinos than were produced by the original production process. Consequently, this would skew the information about the original neutrino source that we would like to extract from observing neutrinos in experiments.

\begin{figure} [htbp]
\centering
    \subfloat[][]{\includegraphics[scale=0.375]{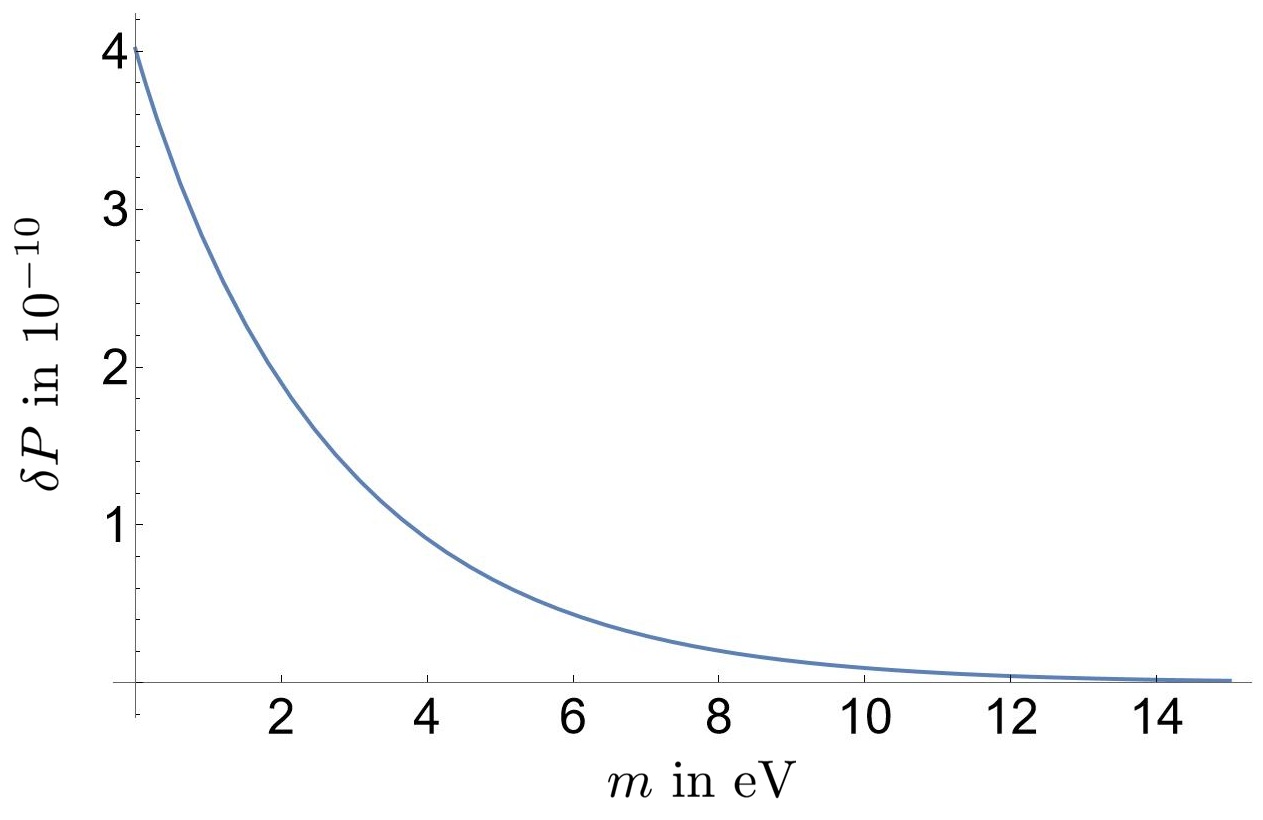}}
    \subfloat[][]{\includegraphics[scale=0.375]{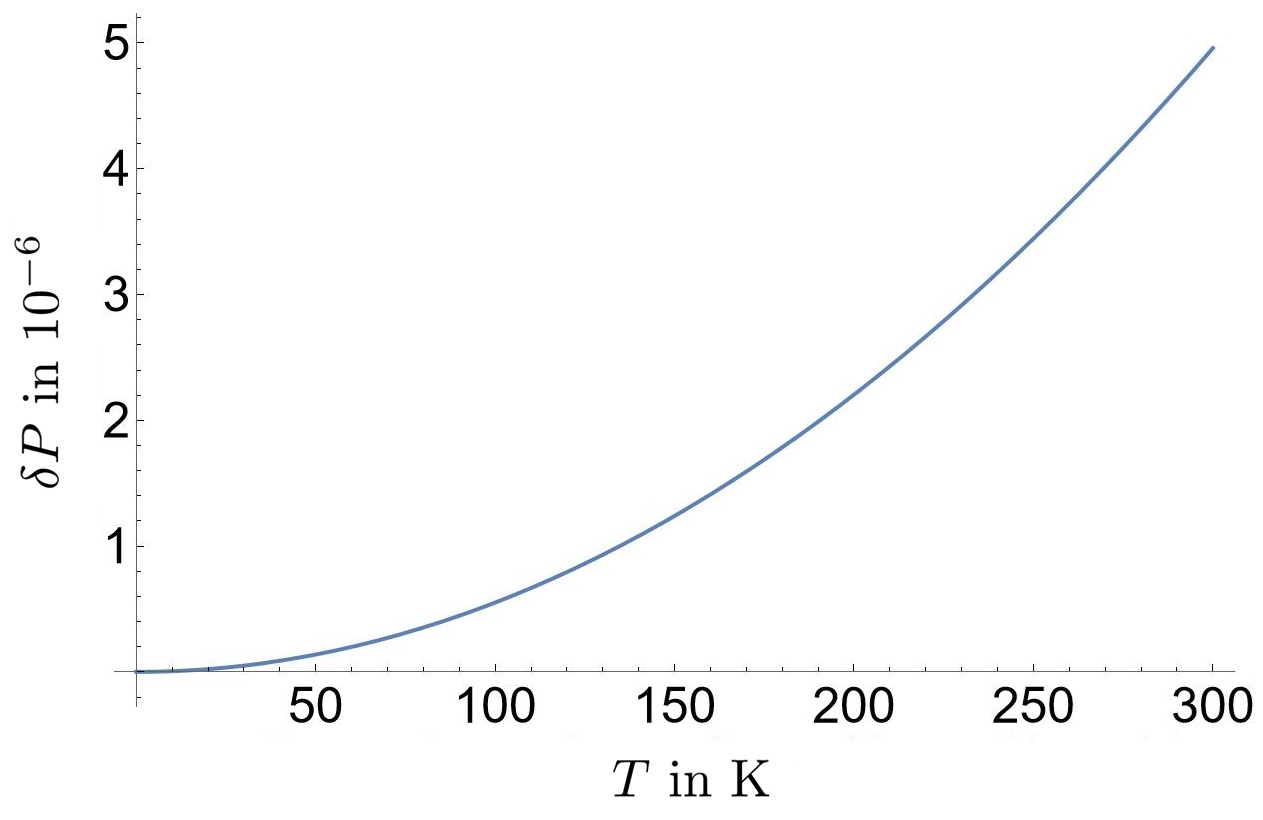}}
    \qquad
    \subfloat[][]{\includegraphics[scale=0.375]{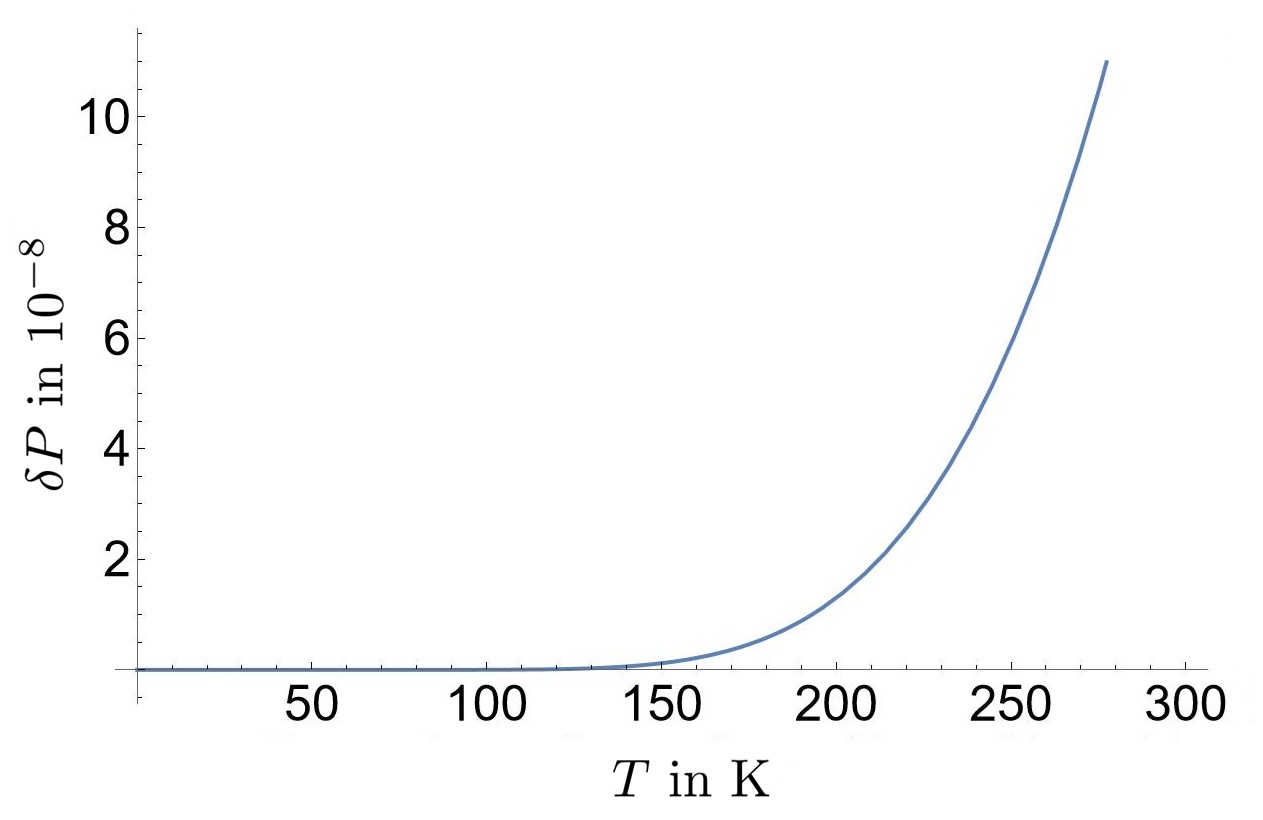}}
    \subfloat[][]{\includegraphics[scale=0.375]{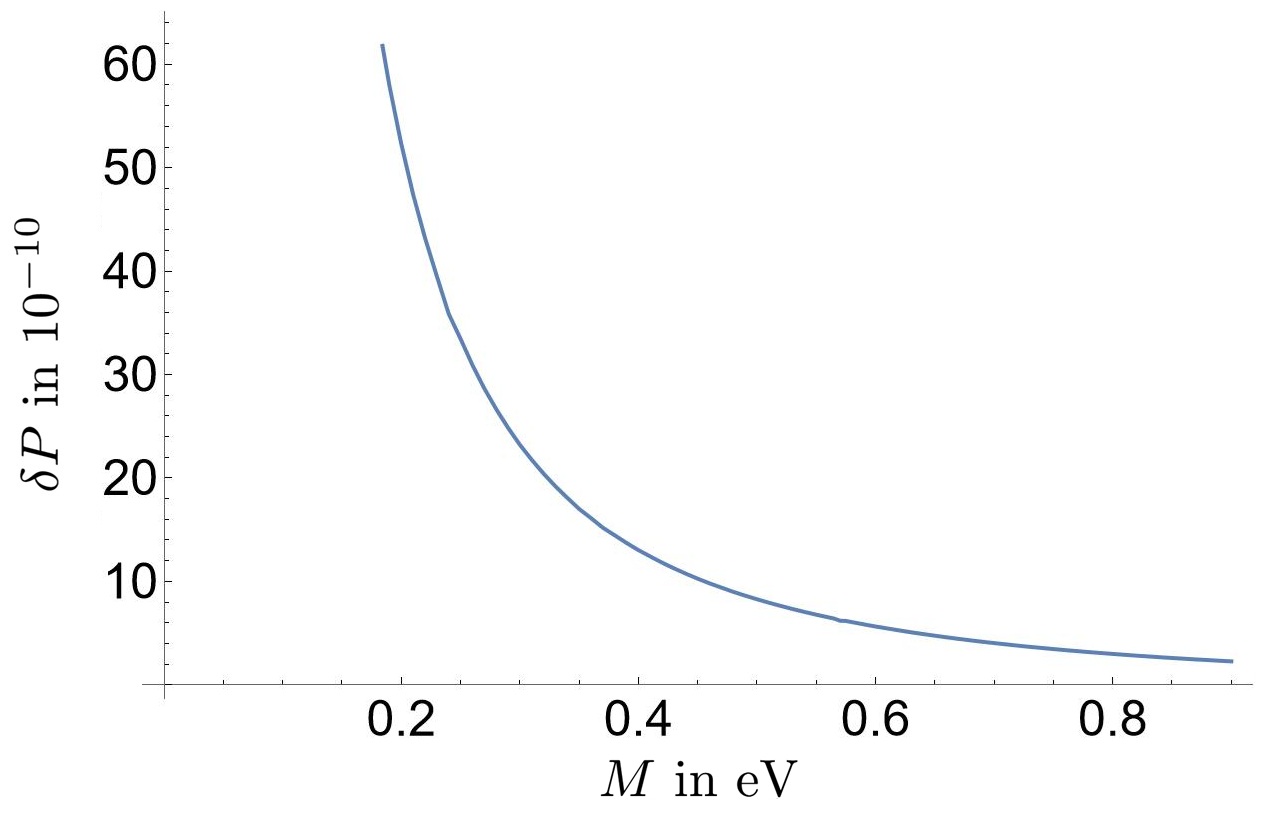}}
\caption{Probability deviation $\delta P$ as defined in Eq.~(\ref{eq:delta}) for the parameters $\lambda = 0.1$ and $t = 1$\,s; (a): over the environment mass $m$ for $M= 0.7$\,eV and $T= 2.7$ K; (b): over the environment temperature $T$ for $M= 0.7$\,eV and $m = 0$; (c): over the environment temperature $T$ for $M= 0.7$\,eV and $m = 0.1$\,eV; (d): over the system mass $M$ for $m = 0$ and $T= 2.7$ K}
    \label{fig:Plots}
\end{figure}

Certainly, the results vary quantitatively and qualitatively with different choices of the model parameters. It is obvious, from Eq.~(\ref{eq:delta}) that the probability deviation changes linearly with $\lambda$. Furthermore, choosing the phase $\theta = \pi$ would flip the sign of $\delta P$ (see the green line in Fig.~\ref{fig:First}) and, therefore, increase the probability of observing the vacuum, while choosing $\theta = \pi/2$ would lead to the deviation oscillating over time with decreasing magnitude around $0$, such that we would effectively not see any changes in probabilities at sufficiently late times; see the orange line in Fig.~\ref{fig:First}. We can also vary the values of $m$ or $T$. Though, while doing so, we must ensure that the product of $\lambda$ and $\Delta^{\mathrm{F}(T\neq 0)}_{zz}$ stays small enough to not violate perturbativity. If we keep $\theta =0$, $T= 2.7$ K, and $\lambda = 0.1$, and choose $t = 1$\,s, then we find that the probability deviation decreases with increasing $m$ ; see Fig.~\ref{fig:Plots}(a). In fact, once the environmental mass becomes significantly larger than the neutrino mass, $\delta P$ becomes negligible, which renders the environment irrelevant to the dynamics of the system. Instead keeping $m=0$ and varying $T$, we can infer from Fig.~\ref{fig:Plots}(b) that $\delta P$ grows rapidly with $T$. In addition, in Fig.~\ref{fig:Plots}(c), we can observe that a higher value of the  mass of the environmental scalar, in this case $m=0.1$\,eV, leads to a less steep growth of $\delta P$ with $T$. For values of $T$ larger than shown, $\delta P$ continues to grow in the same manner, which is not in any way changed by $T$ surpassing the values of $m$ or $M$. Finally, we can also investigate what happens if we consider a different  mass for the system scalar field $M$ while keeping $m=0$ and $T=2.7$\,K. In Fig.~\ref{fig:Plots}(d) we see that $\delta P$ decreases with increasing $M$ and, as expected, diverges for $M \to 0$. This means that increasing either of the two mass scales $M$ or $m$ leads to a lower value of the probability deviation. On the other hand, since our chosen value for $M$ is close to the upper limit for the electron neutrino mass, there is a realistic chance for real neutrinos actually being lighter than considered here. In this case, we can conclude from Fig.~\ref{fig:Plots}(d) that the probability deviation would be significantly larger than what we found in Eq.~(\ref{eq:deltafirst}) and could, therefore, be of observational and experimental significance.


\section{Conclusion}
\label{sec:Conclusion}

The theory of open quantum systems provides us with powerful tools to effectively describe the dynamics of quantum systems interacting with their environments. However, quantum master equations, which are usually used to compute reduced density matrices of open systems, are often too complicated for practicable descriptions of particle number-changing processes without major approximations and simplifications. For this reason, in this article, we have demonstrated how the first principle-based method developed in Ref.~\cite{Kading:2022jjl} can be employed to circumvent such problems in the context of scalar quantum field theory. We have studied a real scalar field $\phi$ as an open quantum system interacting with an environment consisting of a real scalar field $\chi$ with a temperature $T$, focusing on how the probabilities for observing the vacuum or two-particle states evolve with time if those states are initially correlated. We have quantified how the probability of these Fock states change for light scalars, where the environmental temperature is that of the cosmological background today. Our findings suggest that the qualitative and quantitative changes of the probabilities for observing zero- or two-particle neutrino states vary greatly with the particular choices for the initial states and model parameters, which are both strongly dependent on the environment model and the specific process that originally produced neutrinos at the initial time. What makes our results especially interesting is the fact that we have shown that, in principle, it is possible for interactions between neutrinos and an environment to lead to more or less observable neutrinos than those that were originally produced at the source. This may be particularly significant considering that we have used a rather high permitted value for the neutrino mass, which in reality likely will be much smaller, and we have seen that the changes of the probabilities due to the neutrino-environment interactions grow rapidly with a decreasing neutrino mass. Though, it should be stressed that, in a realistic scenario, the produced neutrinos may have varying initial phases $\theta$. As we can see from Fig.~\ref{fig:First}, this would likely lead to a washing out of the effect. Consequently, we expect the most significant reduction of observable neutrinos for sources that result in only a small spread in the phases.

While our results are based on a toy model and, therefore, are not necessarily applicable to the real world they give us a general idea of how interactions between open systems and environments can influence the probabilities for observing certain $N$-particle states. Once the method from Ref.~\cite{Kading:2022jjl} has been extended to also describe spin-$1/2$-particles, we will be able to revisit our discussion and make  predictions for real neutrinos and other fermions. At this point, there will opportunities for a large number of other applications, for example, for describing the production of gravitational waves by decays of heavy particles as in Ref.~\cite{Landini:2025jgj}.


\begin{acknowledgments}
The authors are grateful to P.~Millington for helpful discussions.
This research was funded in whole or in part by the Austrian Science Fund (FWF) [10.55776/PAT8564023], and is based upon work from COST Action COSMIC WISPers CA21106, supported by COST (European Cooperation in Science and Technology). CB is supported by  STFC Consolidated Grant [Grant No. ST/T000732/1]. For open access purposes, the author has applied a CC BY public copyright license to any author accepted manuscript version arising from this submission.
\end{acknowledgments}

\subsection*{Data Availability Statement} In case of legitimate interest, the authors will provide the Mathematica files used for producing the plots in Figs.~\ref{fig:First} and \ref{fig:Plots} on request.

\bibliography{Bib}
\bibliographystyle{JHEP}

\end{document}